\documentclass[prl,twocolumn,preprintnumbers,amsmath,amssymb,floats,citeautoscript,nobalancelastpage,superscriptaddress,showpacs]{revtex4}

\usepackage{graphicx}
\usepackage{dcolumn}
\usepackage{bm}
\usepackage{times}
\usepackage{txfonts}
\usepackage{color}

\begin{document}
	
	\title{Enhanced thermopower via spin-state modification}
	
	\author{Hidefumi~Takahashi}
	\affiliation{Department of Physics, Nagoya University, Nagoya 464-8602, Japan}
	\affiliation{Department of Applied Physics and Quantum Phase Electronics Center (QPEC), University of Tokyo, Tokyo 113-8656, Japan}
	\author{Shintaro~Ishiwata}
	\affiliation{Department of Applied Physics and Quantum Phase Electronics Center (QPEC), University of Tokyo, Tokyo 113-8656, Japan}
	\affiliation{PRESTO, Japan Science and Technology Agency, Kawaguchi, Saitama 332-0012, Japan}
	\author{Ryuji~Okazaki}
	\affiliation{Department of Physics, Faculty of Science and Technology, Tokyo University of Science, Noda, Chiba, 278-8510, Japan}
	\author{Yukio~Yasui}
	\affiliation{Department of Physics, Meiji University, Kawasaki 214-8571, Japan}
	\author{Ichiro~Terasaki}
	\affiliation{Department of Physics, Nagoya University, Nagoya 464-8602, Japan}

	\begin{abstract}
		We investigated the effect of pressure on the magnetic and thermoelectric properties of Sr$_{3.1}$Y$_{0.9}$Co$_{4}$O$_{10+\delta }$. The magnetization is reduced with the application of pressure, reflecting the spin-state modification of the Co$^{3+}$ ions into the nonmagnetic low-spin state. Accordingly, with increasing pressure, the Seebeck coefficient is enhanced, especially at low temperatures, at which the effect of pressure on the spin state becomes significant. These results indicate that the spin-orbital entropy is a key valuable for the thermoelectric properties of the strongly correlated cobalt oxides.   
	\end{abstract}
	
	
	\maketitle
	Transition metal oxides possess a variety of functional properties emerging from the interplay between the charge, spin, and orbital degrees of freedom.\cite{Mott,Imada,Tokura} The giant thermoelectric effect in layered cobalt oxides is considered a typical example. The thermoelectric effect reflects the relation between the electric current density \mbox{\boldmath$j$} and thermal current density \mbox{\boldmath$q$}, which is expressed as \mbox{\boldmath$q$}/$T=\alpha\mbox{\boldmath$j$}$, where $\alpha$ is the Seebeck coefficient. Here \mbox{\boldmath$q$}/$T$ can be regarded as the entropy current density, i.e., $\alpha$ corresponds to the entropy per carrier.\cite{Callen}  In the case of conventional metals, the value of $\alpha$ is small, i.e., of the order of a few $\muup$VK$^{-1}$, indicating the small entropy current carried by the charge degrees of freedom. On the other hand, Na$_{x}$CoO$_{2}$ has an unusually large $\alpha$ and exhibits metallic conductivity, implying an additional contribution to the entropy current other than the charge degrees of freedom.\cite{10}
	
	One possible origin of the giant thermoelectric effect in Na$_{x}$CoO$_{2}$ is the large spin-orbital entropy inherent to the various spin states of Co$^{3+}$ (Co$^{4+}$) ions,\cite{10,Koshibae1,Koshibae2} i.e., high-spin (HS), intermediate-spin (IS), and low-spin (LS) states with the electron configurations, $e_{\rm g}^{2}t_{\rm 2g}^{4}$ ($e_{\rm g}^{2}t_{\rm 2g}^{3}$), $e_{\rm g}^{1}t_{\rm 2g}^{5}$ ($e_{\rm g}^{1}t_{\rm 2g}^{4}$), $e_{\rm g}^{0}t_{\rm 2g}^{6}$ ($e_{\rm g}^{0}t_{\rm 2g}^{5}$), respectively. Thus far, the spin entropy contribution has been suggested as the likely origin of the large value of $\alpha$ of Na$_{x}$CoO$_{2}$, based on the magnetic field dependence of $\alpha$.\cite{Wang} However, the orbital contribution enriched by the spin-state degrees of freedom has not been experimentally clarified, largely because of the difficulty with the manipulation of the spin state. In addition, the large $\alpha$ of Na$_{x}$CoO$_{2}$ can be explained by the unique band structure known as the ``Pudding Mold" type, instead of considering the degeneracy of 3$d$ orbitals. Therefore, the contribution of the spin-orbital entropy to $\alpha$ remains elusive.\cite{12,13}
	
	The A-site ordered perovskite Sr$_{3.1}$Y$_{0.9}$Co$_{4}$O$_{10+\delta}$ is a promising candidate to clarify the relation between the spin-orbital entropy and $\alpha$, because this compound contains Co$^{4+}$ ions yielding hole carriers in the matrix of Co$^{3+}$ ions with various spin states, which can be sensitively controlled by external pressure \cite{1,5,matsunaga}. The crystal structure of Sr$_{3.1}$Y$_{0.9}$Co$_{4}$O$_{10+\delta}$ is shown in Fig. 1. The structure consists of octahedral CoO$_{6}$ layers and tetrahedral/pyramidal CoO$_{4.25}$ layers that are stacked alternately.\cite{4,3,Withers,6} X-ray diffraction, neutron diffraction, and resonant X-ray scattering (RXS) studies showed that both the HS and IS states exist in the CoO$_{4.25}$ and CoO$_{6}$ layers.\cite{19,18,7} The high-pressure neutron diffraction experiments indicate that the spin configuration in the CoO$_{6}$ layer changes into the LS state owing to the pressure-induced enlargement of the crystal-field splitting $\Delta_{\rm CF}$, which separates the $t_{\rm 2g}$ and $e_{\rm g}$ orbitals (see Fig. 1).\cite{18} In this study, we demonstrate the important role of the spin-orbital entropy in the thermoelectric property of cobalt oxides by investigating the effect of pressure on the magnetic and thermoelectric properties of Sr$_{3.1}$Y$_{0.9}$Co$_{4}$O$_{10+\delta}$.
	
	\begin{figure}[t]
		\begin{center}
			\includegraphics[width=6cm]{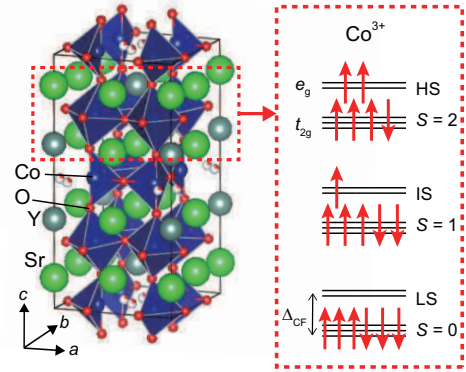}
			\caption{(color online) Crystal structure of Sr$_{3.1}$Y$_{0.9}$Co$_{4}$O$_{10+\delta}$. Octahedral CoO$_{6}$ and tetrahedral/pyramidal CoO$_{4.25}$ layers are alternately stacked with insertion of the ordered Sr$_{0.75}$Y$_{0.25}$O layer. The high-spin (HS) and/or intermediate-spin (IS) states of the Co$^{3+}$ in the CoO$_{6}$ layer (enclosed within the red dashed square) at ambient pressure change into the low-spin (LS) state with the application of physical pressure.}
		\end{center}
	\end{figure}
	
	\begin{figure}[t]
		\begin{center}
			\includegraphics[width=7cm]{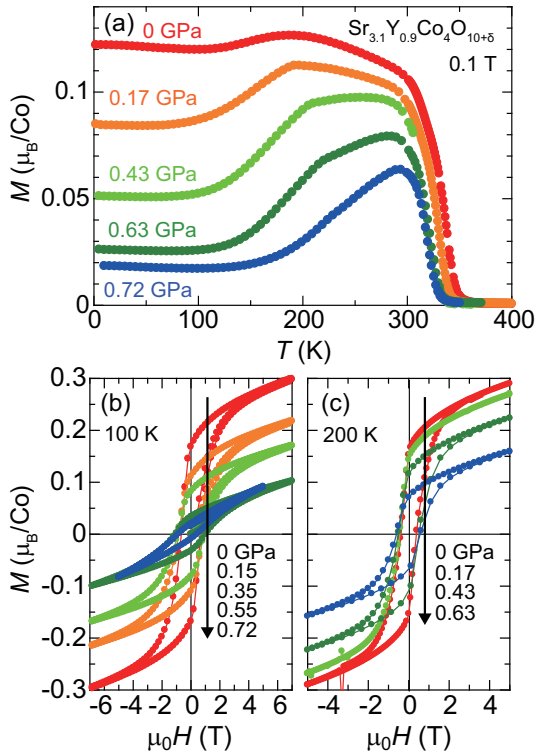}
			\caption{(color online) (a) Temperature dependence of the magnetic susceptibility for various pressures at 0.1 T. Magnetic field dependence of the magnetization for various pressures at 100 K (b) and 200 K (c).}
		\end{center}
	\end{figure}
	
	\begin{figure}[htbp]
		\begin{center}
			\includegraphics[width=7cm]{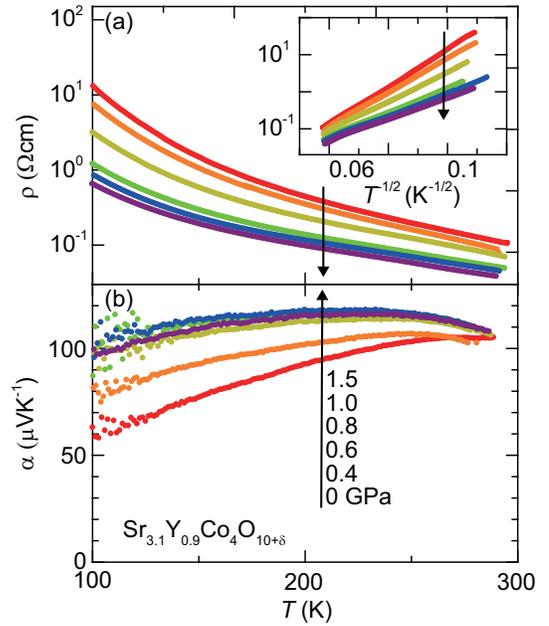}
			\caption{(color online) Temperature dependence of the resistivity $\rho$ (a) and Seebeck coefficient $\alpha$ (b). The inset in Fig. 3(a) shows $\rho$ as a function of $T^{-1/2}$, indicating variable-range-hopping (VRH) conduction with large Coulomb interaction.}
		\end{center}
	\end{figure}
	
	\begin{figure}[htbp]
		\begin{center}
			\includegraphics[clip,width=8cm]{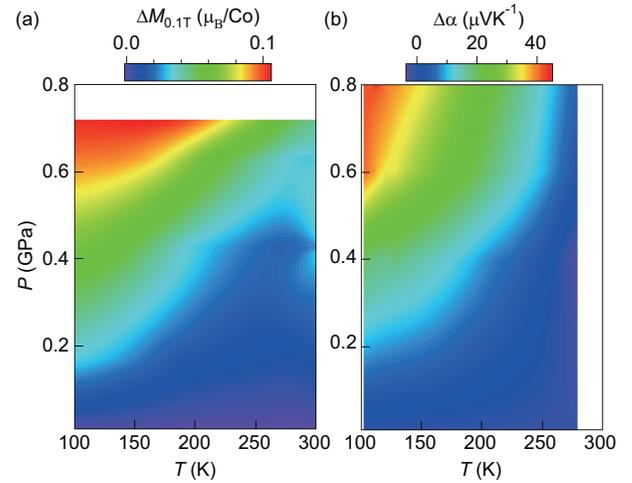}
			\caption{(color online) Contour plots of $\Delta M_{0.1T}$ [$=M$(0 GPa)-$M$($P$) at 0.1T] (a) and $\Delta \alpha$ [$=\alpha$($P$)-$\alpha$(0 GPa)] (b) as functions of temperature $T$ and pressure $P$.}
		\end{center}
	\end{figure} 
	
	\begin{figure}[htbp]
		\begin{center}
			\includegraphics[clip,width=8cm]{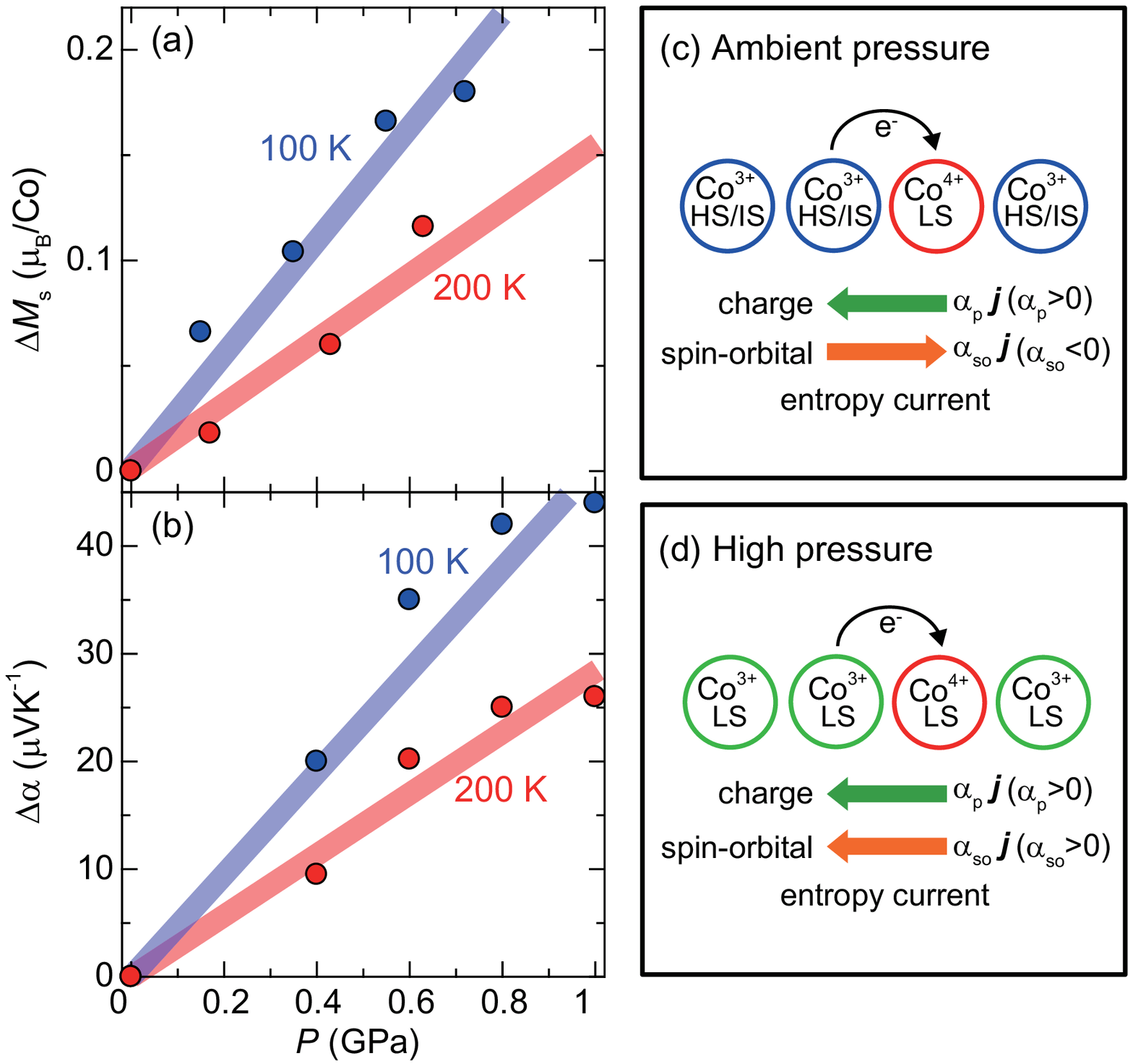}
			\caption{(color online) (a) Pressure dependence of $\Delta M_{\rm s}$ [$= M_{\rm s} (0)- M_{\rm s} (P)$] at 100 and 200 K, where the saturated magnetization $M_{\rm s}$ is estimated by extrapolation of the linear part of the $M-H$ curves. (b) Increase in the Seebeck coefficient $\Delta \alpha$ as a function of pressure at 100 and 200 K. Schematic illustrations of the carrier and entropy currents at ambient (c) and high pressures (d).}
		\end{center}
	\end{figure} 
	
	A polycrystalline sample of Sr$_{3.1}$Y$_{0.9}$Co$_{4}$O$_{10+\delta }$ was prepared by a solid-state reaction method. A mixture of SrCO$_{3}$, Y$_{2}$O$_{3}$, and Co$_{3}$O$_{4}$ with 5-mol $\%$ excess Co with regard to the nominal composition was calcined at 1100 $^{\circ}$C for 12 h in air. Note that samples prepared from a stoichiometric mixture included a tiny fraction of unreacted Y$_{2}$O$_{3}.$\cite{5} The calcined product was ground, pressed into a pellet, and sintered at 1100 $^{\circ}$C for 48 h in air. The magnetization-temperature and magnetization-field curves were measured by a commercial superconducting quantum interference device magnetometer (Quantum Design MPMS). The temperature dependence of the resistivity $\rho$ and $\alpha$ was measured in the temperature range from 290 to 80 K by a conventional four probe method and a steady-state method, respectively. The external pressure was applied at room temperature using a piston cylinder clamp cell made of CuBe. Daphne oil 7373 was used as the pressure-transmitting medium. 
	
	We verify the spin-state modification in Sr$_{3.1}$Y$_{0.9}$Co$_{4}$O$_{10+\delta}$ from the pressure dependence of magnetization $M$, shown in Fig. 2(a). At ambient pressure, a ferro(ferri-)magnetic transition occurs at 340 K (Fig. 2(a)); the spins in the CoO$_{4.25}$ layers are antiferromagnetically  aligned with the HS state, and the existence of both the HS and IS states in CoO$_{6}$ layers is considered to be responsible for the ferrimagnetism.\cite{1,19,7} In addition, the weak anomaly observed at approximately 180 K is attributed to the spin-state modification of Co$^{3+}$; a small portion of the HS or IS state of the Co$^{3+}$ ions is converted into the nonmagnetic LS state with decreasing temperature as well as LaCoO$_{3}$.\cite{kimura,Raccah,Korotin} At low temperatures, $M$ dramatically decreases upon the application of pressure up to 1 GPa. Figures 2(b) and 2(c) show the magnetic field dependence of $M$ at various pressures at 100 K and 200 K, respectively. The saturated magnetization $M_{\rm s}$ is suppressed with increasing pressure, indicating that the spin state of the Co$^{3+}$ ions changes from the HS or IS state with $S$ = 2 or 1 to the LS state with $S$ = 0.   
	
	Figures 3(a) and (b) show the temperature dependence of $\rho$ and $\alpha $ at high pressures up to 1.5 GPa. The value of $\alpha $, which has a positive sign, gradually decreases in a concave upward manner on decreasing temperature, as typically seen in a disordered semiconductor.\cite{Burns} Likewise, $\rho$ displays nonmetallic behavior, and log$\rho$ is linear to $T^{-1/2}$ as shown in the inset in Fig. 3(a), indicating that the carrier conduction can be described by variable-range-hopping (VRH) with large Coulomb interactions.\cite{16,17} These results imply that the incoherent hopping conduction is dominated by a small number of holes on the minority Co$^{4+}$ ions generated by the excess oxygen, which conduct in the octahedral CoO$_{6}$ layers as incoherent hopping carriers.\cite{1}
	
	As the pressure increases from 0 to 1.5 GPa, $\rho$ is reduced by one order of magnitude with retention of the nonmetallic behavior. For transition metal oxides, resistivity reduction with increasing pressure has been observed as a result of bandwidth broadening. For example in La$_{0.9}$Sr$_{0.1}$CoO$_{3}$, the resistivity of $\sim$ 0.1 $\Omega$ cm at 300 K at ambient pressure decreases by a factor of 10 at 2 GPa.\cite{Mydeen} Thus, it is presumable that the same mechanism works for the pressure-induced resistivity reduction in the present system. On the other hand, upon increasing the pressure up to 0.8 GPa, at low temperatures $\alpha$ is dramatically enhanced from 50 to 100 $\muup$VK$^{-1}$, followed by saturation above 1 GPa. The observation of the strong enhancement of $\alpha$ is incompatible with the classical band model where $\alpha$ is roughly given as a function of the number of holes per site, $p$, as $-$($k_{\rm B}$ln$p$)/$e$ for $p\ll 1$.\cite{Zaiman,5} If the enhancement of $\alpha$ ($\sim 50$ $\muup$VK$^{-1}$) is dominated by $p$, the application of pressure should reduce $p$ to nearly half of its original value, which results in an increase in $\rho$. However, $\rho$ decreases with increasing pressure in our experiment, suggesting that $p$ does not determine the value of $\alpha$. Furthermore, the variation in $\alpha$ is remarkable compared with other strongly correlated electron systems such as the analogous transition-metal oxide La$_{1-x}$Sr$_{x}$CuO$_{4}$ ($x=0.22$) and the heavy electron materials YbCu$_{2}$Si$_{2}$, which exhibit substantially small pressure effects of 5 $\muup$VK$^{-1}$/GPa and $-3$ $\muup$VK$^{-1}$/GPa at 100 K through a band structure modification and weakening of the Kondo effect, respectively.\cite{14,15}
	
	In order to clarify the origin of the unusual enhancement of $\alpha$, we compare the effect of pressure on $M$ with that on $\alpha$. The counter plots of $\Delta M_{0.1T}$ [$=M$(0 GPa)$-M$($P$) at 0.1T] and $\Delta \alpha$ [$=\alpha$($P$)$-\alpha$(0 GPa)] as functions of temperature $T$ and pressure $P$ are depicted in Fig. 4(a) and 4(b), respectively. It is noted here that both plots show highly similar features, suggesting the close relation between the suppression of $M$ through the spin-state modification and the highly enhanced $\alpha$.   
	
	Here, we have presented further insight into the effect of pressure on $M$ and $\alpha$ to clarify the impact of the spin-state modification on $\alpha$. Figure 5(a) shows $\Delta M_{\rm s}= M_{\rm s} (0)- M_{\rm s} (P)$ as a function of pressure, where the saturated magnetic moment $M_{\rm s}$ is evaluated from the $M-H$ curve. 	At 100 and 200 K, $\Delta M_{\rm s}$ linearly decreases with increasing pressure at the rate of 0.3 ($\mu _{\rm B}$/Co)/GPa and 0.15 ($\mu _{\rm B}$/Co)/GPa, respectively.Similar linearity in the pressure dependence of $\Delta \alpha=\alpha (P)-\alpha (0)$ is observed at 100 and 200 K as shown in Fig. 5(b), where the application of pressure enhances  $\Delta \alpha$. The linear pressure dependence of $\Delta M_{\rm s}$ and $\Delta \alpha$ enables us to evaluate the proportional coefficient $\gamma $ determined by $\Delta \alpha (P)= \gamma \Delta M_{\rm s} (P)$ below 1 GPa. Interestingly, $\gamma$ is found to have almost the same value of 200 ($\muup$V/K)/($\mu _{\rm B}$/Co) at 100 and 200 K, demonstrating that the suppression of the magnetic moment enhances $\alpha$.
	
	Considering the close relation between the pressure dependence of $\Delta M_{\rm s}$ and $\Delta \alpha$, it is reasonable that the spin-state modification, which changes the spin-orbital entropy, affects the value of $\alpha$. Given that the transport properties are characterized by the incoherent hopping of the localized charge carriers, which are the holes of the Co$^{4+}$ in the matrix of the Co$^{3+}$ for Sr$_{3.1}$Y$_{0.9}$Co$_{4}$O$_{10+\delta}$, $\alpha$ is linked to the spin-orbital entropy by means of the extended Heikes formula:\cite{Koshibae1} 
	\begin{equation}
	\alpha  (T)=-\frac{k_{\rm B}}{e}\ln\left[\frac{g_{3}}{g_{4}}\frac{p}{1-p}\right],
	\label{eq:s}
	\end{equation} 
	where $g_{3}$ and $g_{4}$ are the spin-orbital degeneracy of the Co$^{3+}$ and Co$^{4+}$ ions, respectively, and $p$ is the concentration of the Co$^{4+}$ ion corresponding to the hole density.This expression can be divided into the terms representing the carrier concentration $\alpha_{\rm p}$ ($=-(k_{\rm B}/e)$ln[$p$/($1-p$)]) and the spin-orbital degeneracy(entropy) $\alpha_{\rm so}$ ($=-(k_{\rm B}/e)$ln[$g_{3}$/$g_{4}$]).
	
	Here, we estimate the contribution of the spin-orbital entropy to $\alpha$ with this extended Heikes formula.
	Note that above $T_{\rm N}$ the Heikes formula is no longer appropriate to evaluate $\alpha$, because the system becomes metallic and $\alpha$ becomes considerably small.\cite{HT_SYCO} On the other hand, since thermopower around room temperature is severely affected by spin fluctuation enhanced around $T_{\rm N}$, we have conducted the analysis based on the Heikes formula at low temperature well below $T_{\rm N}$. It has been reported that this formula is applicable to obtain a rough estimation for $\alpha$ below room temperature in a system with correlated hopping transports,\cite{Taskin} which is also the case for the present system showing nonmetallic behavior even at high pressures. In addition, when pressure is applied, $p$ seems to be constant in this system, since the carrier transport arises from the hopping conduction of the holes on the Co$^{4+}$ ions, which result from the oxygen nonstoichiometry. We, therefore, expect that only $\alpha_{\rm so}$ is substantially enhanced by the pressure.
	
	For Sr$_{3.1}$Y$_{0.9}$Co$_{4}$O$_{10+\delta}$, $\alpha _{\rm so}$ is determined by the spin state of the CoO$_{6}$ layers, in which the HS and/or IS state exist at ambient pressure. As has been reported in previous studies involving neutron scattering measurements, the pressure induces the spin-state modification, where half of the spins in the CoO$_{6}$ layers changes to the LS state at 2 GPa.\cite{18} Provided that all the HS or IS states of Co$^{3+}$ change to the LS state, $\Delta \alpha_{\rm so}$ [$=\alpha_{\rm so}$(LS)-$\alpha_{\rm so}$(HS) or $\alpha_{\rm so}$(IS)] is evaluated to be ($k_{\rm B}/e$)ln(15/1)$\sim230$ $\muup$VK$^{-1}$(HS) or ($k_{\rm B}/e$)ln(18/1)$\sim250$ $\muup$VK$^{-1}$(IS), where the LS state of the Co$^{4+}$ is maintained under pressure: the values of $g_{3}$ of the HS, IS, and LS states are 15, 18, and 1, respectively, and $g_{4}$ of the LS state is 6.\cite{Koshibae1}	Given that 25 \% of the Co$^{3+}$ ions in the CoO$_{\rm 6}$ layers changes to the LS state at 1 GPa as expected from the neutron scattering measurements, $\Delta \alpha_{\rm so}$ is evaluated to be $\sim60$ $\muup$V/K, which is comparable to the experimental value at 100 K ($\sim$ 50 $\muup$V/K). 
	
	Finally, we analyze the effect of pressure on $\alpha$ in terms of the charge and entropy current. At ambient pressure, Co$^{3+}$ and Co$^{4+}$ are in the HS/IS and LS states, respectively. The HS/IS (Co$^{3+}$) commonly possesses higher degeneracy than the LS (Co$^{4+}$) due to the larger spin and orbital degrees of freedom. This situation yields a negative value of $\alpha_{\rm so}$, since the value of $g_{3}/g_{4}$ exceeds unity. On the other hand, when the HS/IS (Co$^{3+}$) is converted into the LS state at high pressure, the value of $g_{3}/g_{4}$ becomes smaller than unity, since the spin and orbital degeneracy of the LS state disappears. In this case, the value of $\alpha_{\rm so}$ should be positive. While the observed large positive $\alpha$ is attributed to the carrier contribution $\alpha_{\rm p}$ because of the small carrier concentration ($p\ll 1$),\cite{5} the putative sign change of $\alpha_{\rm so}$ reflecting the spin state modification can be a source of the positive enhancement of $\alpha$ $(=\alpha_{\rm p}+\alpha_{\rm so})$. In other words, at ambient pressure, the entropy current associated with the hopping of hole carriers of Co$^{4+}$ $\alpha_{\rm p}$\mbox{\boldmath$j$} in the matrix of HS/IS Co$^{3+}$ is counterbalanced by the spin-orbital entropy current $\alpha_{\rm so}$\mbox{\boldmath$j$} ($\alpha_{\rm p}>0$, $\alpha_{\rm so}<0$). In contrast, at high pressure, for the spin-state modification of Co$^{3+}$ to the LS state, both entropies are additive to the entropy current ($\alpha_{\rm p}>0$, $\alpha_{\rm so}>0$), resulting in the positive enhancement of $\alpha$ (Fig. 5). This result proposes that, in transition-metal compounds, the manipulation of the spin and orbital by changing the lattice parameters with the isovalent chemical substitution can significantly improve the thermoelectric efficiency.\cite{5}
	
	In summary, we measured the pressure dependence of the magnetization, electrical resistivity, and Seebeck coefficient for Sr$_{3.1}$Y$_{0.9}$Co$_{4}$O$_{10+\delta}$. The Seebeck coefficient undergoes significant enhancement relevant to the suppression of the magnetic moment, which is induced by the spin-state modification of the Co$^{3+}$ spins in the CoO$_{6}$ layers. Our results revealed that the thermoelectric properties are subject to the spin-orbital entropy, and stimulate the exploration of new thermoelectric materials by manipulating the spin state of a magnetic compound.

	\section{Acknowledgements}
	This work was partially supported by KAKENHI (Grants No. 17H01195 and No. 16K17736), JST PRESTO Hyper-nano-space design toward Innovative Functionality (JPMJPR1412), the Asahi Grass Foundation, and the Murata Science Foundation.


\begin{thebibliography}{99}
		\bibitem{Mott} N. F. Mott, $Metal-Insulator$ $Transitions$ (Taylor and Francis, London,1990).
		\bibitem{Imada} M. Imada, A. Fujimori, and Y. Tokura, Rev. Mod. Phys. \textbf{70}, 1039 (1998).
		\bibitem{Tokura} $Colossal$ $Magnetoresistive$ $Oxides$, edited by Y. Tokura (Gordon and Breach Science Publishers, New York, 2000).
		\bibitem{Callen} H. B. Callen, $Thermodynamics$ $and$ $introduction$ $to$ $thermostatistics$ (Second Edn.) (Wiley, New York, 1985).
		\bibitem{10} I. Terasaki, Y. Sasago, and K. Uchinokura, Phys. Rev. B \textbf{56,} R12685(R) (1997).  
		\bibitem{Koshibae1} W. Koshibae, K. Tsutsui, and S. Maekawa, Phys. Rev. B \textbf{62,} 6869 (2000).
		\bibitem{Koshibae2} W. Koshibae and S. Maekawa, Phys. Rev. Lett. \textbf{87,} 236603 (2001).
		\bibitem{Wang} Y. Wang, N. S. Rogado, R. J. Cava, and N. P. Ong, Nature(London) \textbf{423,} 425 (2003). 
		\bibitem{12} D. J. Singh, Phys. Rev. B \textbf{61,} 13397 (2000).
		\bibitem{13} K. Kuroki and R. J. Arita, Phys. Soc. Jpn. \textbf{76,} 083707 (2007).
		\bibitem{1} W. Kobayashi, S. Ishiwata, I. Terasaki, M. Takano, I. Grigoraviciute, H. Yamauchi, and M. Karppinen, Phys. Rev. B \textbf{72,} 104408 (2005).
		\bibitem{5} S. Yoshida, W. Kobayashi, T. Nakano, I. Terasaki, K. Matsubayashi, Y. Uwatoko, I. Grigoraviciute, M. Karppinen, and H. Yamauchi, J. Phys. Soc. Jpn. \textbf{78,} 094711 (2009).
		\bibitem{matsunaga} T. Matsunaga, T. Kida, S. Kimura, M. Hagiwara, S. Yoshida, and I. Terasaki, J. Phys.: Conf. Ser. \textbf{200}, 012116 (2010).  
		\bibitem{Withers} R. L. Withers, M. James, D. J. Goossens, J. Solid State Chem. \textbf{174,} 198 (2003).
		\bibitem{3} S. Ya. Istomin, J. Grins, G. Svensson, O. A. Drozhzhin, V. L. Antipov, and J. P. Attfield, Chem. Mater. \textbf{15,} 4012 (2003).
		\bibitem{4} S. Ya. Istomin, O. A. Drozhzhin, G. Svensson, and E. V. Antipov, Solid State Sci. \textbf{6,} 539 (2004).
		\bibitem{6} D. V. Sheptyakov, V. Yu. Pomjakushin, O. A. Drozhzhin, S. Ya. Istomin, E. V. Antipov, I. A. Bobrikov, and A. M. Balagurov, Phys. Rev. B \textbf{80,} 024409 (2009).
		\bibitem{19} S. Ishiwata, W. Kobayashi, I. Terasaki, K. Kato, and M. Takata, Phys. Rev. B \textbf{75,} 220406(R) (2007).
		\bibitem{18} N. O. Golosova, D. P. Kozlenko, L. S. Dubrovinsky, O. A. Drozhzhin, S. Ya. Istomin, and B. N. Savenko, Phys. Rev. B \textbf{79,} 104431 (2009).
		\bibitem{7} H. Nakao, T. Murata, D. Bizen, Y. Murakami, K. Ohoyama, K. Yamada, S. Ishiwata, W. Kobayashi, and I. Terasaki, J. Phys. Soc. Jpn. \textbf{80,} 023711 (2011).
		\bibitem{kimura} S. Kimura, Y. Maeda, T. Kashiwagi, H. Yamaguchi, M. Hagiwara, S. Yoshida, I. Terasaki, and K. Kindo, Phys. Rev. B \textbf{78,} 180403(R) (2008).
		\bibitem{Raccah} P. M. Raccah and J. B. Goodenough, Phys. Rev. \textbf{155,} 932 (1967).
		\bibitem{Korotin} M. A. Korotin, S. Yu. Ezhov, I. V. Solovyev, V. I. Anisimov, D. I. Khomskii, and G. A. Sawatzky, Phys. Rev. B \textbf{54,} 5309 (1996).
		\bibitem{Burns} M. J. Burns and P. M. Chaikin, Phys. Rev. B \textbf{27,} 5924 (1983).
		\bibitem{16} A. L. Efros and B. I. Shklovskii, J. Phys. C \textbf{8,} L49 (1975).
		\bibitem{17} B. I. Shklovskii and A. L. Efros, $Electronic$ $Properties$ $of$ $Doped$ $Semiconductors$ (Springer-Verlag, Berlin, 1984).
		\bibitem{Mydeen} K. Mydeen, P. Mandal, D. Prabhakaran, and C. Q. Jin, Phys. Rev. B \textbf{80,} 014421 (2009).
		\bibitem{Ziaman} J. M. Ziman, $Principles$ $of$ $The$ $Theory$ $of$ $Solids$ (Cambridge Univ. Press, London, 1972).
		\bibitem{14} J. -S. Zhou and J. B. Goodenough, Phys. Rev. Lett. \textbf{77,} 151 (1996).
		\bibitem{15} K. Alami-Yadri, D. Jaccard, and D. Andreica, J. Low Temp. Phys. \textbf{114,} 135 (1999).
		\bibitem{HT_SYCO} W. Kobayashi, S. Yoshida, and I. Terasaki, J. Phys. Soc. Jpn. \textbf{75}, 103702 (2006).
		\bibitem{Taskin} A. A. Taskin, A. N. Lavrov, and Y. Ando, Phys. Rev. B \textbf{73,} 121101(R) (2006).  
	\end{thebibliography}
\end{document}